\documentclass[reprint,NumberedRefs]{JASA_mod}

\begin{document}
%% the square bracket argument will send term to running head in
%% preprint, or running foot in reprint style.

\title[Frequency-domain sound field as a band-limited function]{Frequency-Domain Sound Field from the Perspective of Band-Limited Functions}

% ie
%\title[JASA/Sample JASA Article]{Sample JASA Article}

%% repeat as needed
\author{Takahiro Iwami}
\email{iwami@design.kyushu-u.ac.jp}
\affiliation{Graduate School of Design, Kyushu University, 4-9-1 Shiobaru, Minamiku, Fukuoka, 815-8540, Japan}

\author{Akira Omoto}
\affiliation{Faculty of Design, Kyushu University, 4-9-1 Shiobaru, Minamiku, Fukuoka, 815-8540, Japan}

%% For preprint only,
%  optional, if you want want this message to appear in upper right corner of title page
% \preprint{}

%ie
%\preprint{Author, JASA}

% optional, if desired:
%\date{\today}

\begin{abstract}
A model that approximates the sound field well is useful in various fields such as acoustic signal processing and numerical simulation.
We have proposed an effective model in which the wideband instantaneous sound field is regarded as an element of a spherically band-limited function space, using the reproducing kernel of that space.
In this paper, the frequency-domain sound field is regarded as an element of some band-limited function space, and a representation of the field as a linear combination of the reproducing kernel in that space is proposed.
This model has the strongest representational capacity of all function systems when we know only the sound pressure information at arbitrary positions.
The proposed model can be considered a generalization of the existing three-dimensional sound field model using the reproducing kernel of the solution space of the Helmholtz equation to the spatial dimension.
One of the advantages of capturing the frequency-domain sound field in this way is the simplicity achieved for the estimation formula of the wavenumber spectrum.
Two numerical simulations were conducted to validate the proposed methods.
\end{abstract}

%% pacs numbers not used

\maketitle

%  End of title page for Preprint option --------------------------------- %

%--------------------------------------%
\section{\label{sec:1} INTRODUCTION}
%--------------------------------------%
The choice of sound field expression is an important factor in various application domains.
For example, in the field of sound field reproduction, wave field synthesis \cite{Berkhout1993, Spors2008} adopts the Rayleigh integral whereas ambisonics \cite{Cooper1972, Gerzon1973, Poletti2005} adopts the spherical harmonic expansion expression.
In addition, numerical calculations are performed in architectural acoustics to evaluate the acoustic properties of materials, such as sound absorption and scattering.
Methods such as the finite element method\cite{Ihlenburg1998} and boundary element method\cite{Marburg2008} are widely used for this purpose. They are based on the weak-form equation and the boundary integral equation, respectively.

Some of the above methods express the sound field in integral form, but in their implementation, all the methods express the sound field as a linear sum of some system of functions $\{ \phi_{n} \}$; i.e., $p = \sum_{n} a_{n} \phi_{n}$.
Most methods assume a linear model in this way, with the exceptions being methods that adopt neural networks\cite{Bishop2006}.
The choice of a system of functions is a difficult problem, especially when there are obstacles in the field; i.e., when the system satisfies an inhomogeneous wave equation.
For example, ambisonics uses spherical harmonics functions, which is a complete orthonormal system of $L^{2}$ on a sphere.
Ueno \textit{et al.}\cite{Ueno2018} constructed a reproducing kernel Hilbert space (RKHS) by appropriately introducing a norm and inner product into the function space consisting of the solution set of the interior problem of the three-dimensional Helmholtz equation, and they applied the reproducing kernel (RK) to a system of functions.

We have so far constructed an RKHS formed by band-limited instantaneous sound fields\cite{Iwami2023, Iwami2024}.
This construction assumes that the instantaneous sound field belongs to a band-limited space such that the wavenumber spectral support is contained in a hypersphere of radius determined by the upper frequency, and we apply the RK of this space to the system of functions.
This method has the advantage that the wavenumber spectrum can be easily estimated by considering the instantaneous sound field as an element of the space of bandwidth-limited functions.

This paper follows previous work and considers the frequency-domain sound field of a general dimension as an element of a space of band-limited functions.
Specifically, we construct a space of band-limited functions such that the wavenumber spectrum exists only on a hyperspherical surface of radius dependent on the frequency under consideration.
The derived RK is equivalent to that derived by Ueno \textit{et al.}, such that the proposed methodology is, in a sense, a generalization of that of Ueno \textit{et al}.
We show that the sound field estimated using this RK is the best approximation for fixed sampling points and that the wavenumber spectrum can be easily estimated from this representation.

Numerical experiments were conducted to verify the validity of the proposed method.

%--------------------------------------%
\section{\label{sec:2} Space of band-limited functions to which frequency-domain sound fields belong and its RK}
%--------------------------------------%
In this paper, the temporal Fourier transform and the $d$-dimensional spatial Fourier transform are defined as
\begin{equation} \label{eq:temporal_FT}
    \hat{f}(\omega) = \mathcal{F}_{\mathrm{t}} f (\omega) := \frac{1}{\sqrt{2\pi}} \int_{\mathbb{R}} f(t) \mathrm{e}^{\mathrm{i} \omega t} \, \mathrm{d} t,
\end{equation}
\begin{equation} \label{eq:spatial_FT}
    F(\bm{k}) = \mathcal{F}_{\mathrm{s}} f (\bm{k}) := \frac{1}{(2\pi)^{\frac{d}{2}}} \int_{\mathbb{R}^{d}} f(\bm{r}) \mathrm{e}^{- \mathrm{i}\bm{k}\cdot\bm{r}} \, \mathrm{d}\bm{r},
\end{equation}
where $t, \omega \in \mathbb{R}$ denotes time and angular frequency and $\bm{r}, \bm{k} \in \mathbb{R}^{d}$ denotes position and wavenumber vectors.

We now consider a sound field in a source-free region in $d$ dimensions.
That is, the sound field $p$ is governed by the $d$-dimensional homogeneous wave equation
\begin{equation}
    \Box p(\bm{r},t) = \Delta p(\bm{r},t) - \frac{1}{c^{2}} \frac{\partial^{2} p}{\partial t^{2}} (\bm{r},t) = 0,
\end{equation}
where $c$ is the speed of sound.
In this case, the wavenumber spectrum can be written as
\begin{equation} \label{eq:wavenumber_spectrum_t}
    P(\bm{k},t) = \overline{P_{\mathrm{b}}(\bm{k})} \mathrm{e}^{\mathrm{i} \omega t} + P_{\mathrm{f}}(\bm{k}) \mathrm{e}^{-\mathrm{i} \omega t},
\end{equation}
where the coefficient function $P_{\mathrm{f}}$ and $\overline{P_{\mathrm{b}}}$ represent the magnitude and initial phase of the plane forward and backward waves concerning the included wavenumber vector $\bm{k}$ in the sense\cite{Iwami2023} that
\begin{align}
    p(\bm{r},t) &= \frac{2}{(2\pi)^{\frac{d}{2}}} \int_{\mathbb{R}^{d}} \mathrm{Re} \left \{ P_{\mathrm{f}}(\bm{k}) \mathrm{e}^{\mathrm{i} (\bm{k}\cdot \bm{r} \mathalpha{-} \omega t)} \right \} \mathrm{d} \bm{k} \notag \\
    &= \frac{2}{(2\pi)^{\frac{d}{2}}} \int_{\mathbb{R}^{d}} \mathrm{Re} \left \{ \overline{P_{\mathrm{b}}(\bm{k})} \mathrm{e}^{\mathrm{i} (\bm{k}\cdot \bm{r} \mathalpha{+} \omega t)} \right \} \mathrm{d} \bm{k}.
\end{align}

\begin{figure}[t]
    \centering
    \includegraphics[width=1.0\columnwidth]{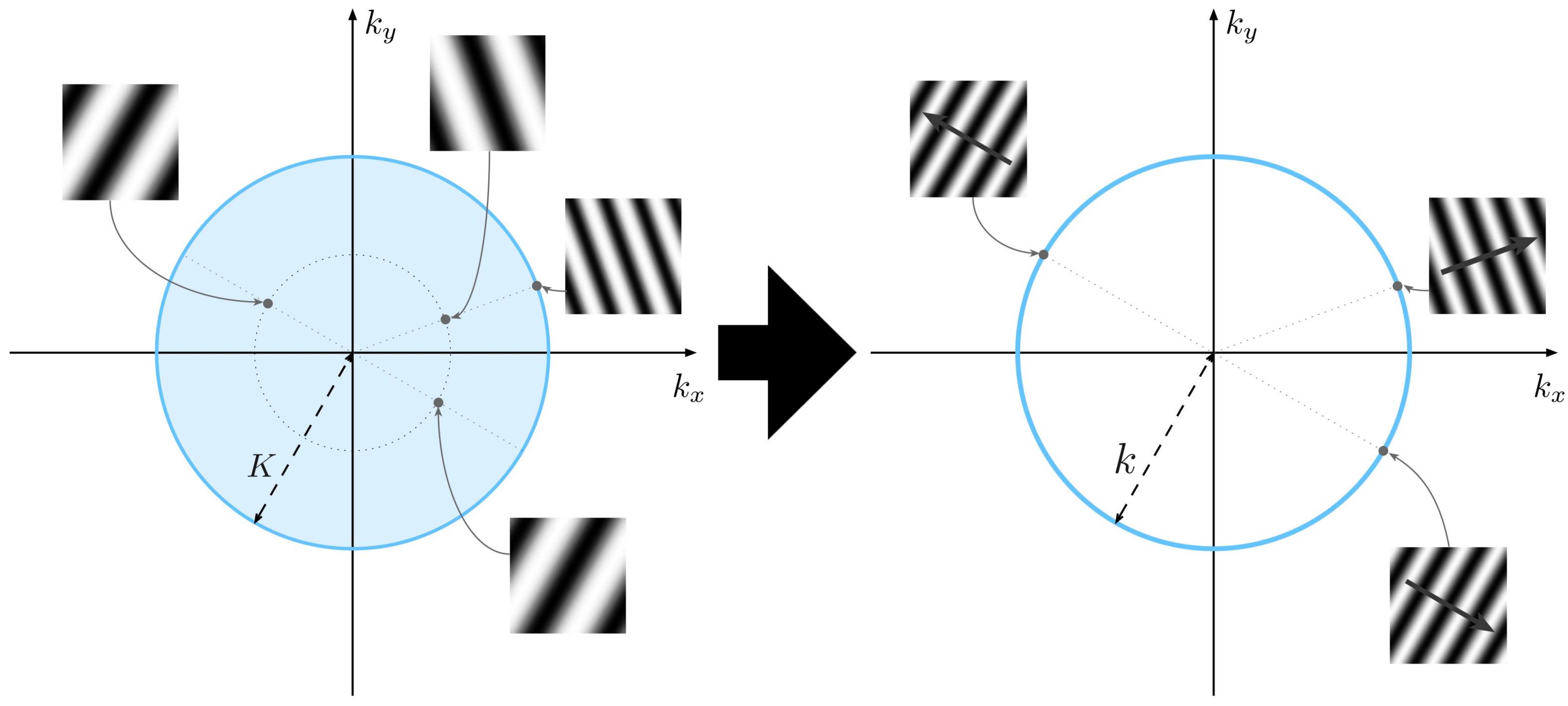} \vspace{0pt}
    \caption{\label{fig1}Visualization of the wavenumber spectrum for the existing model and proposed model. Whereas the existing model can have the entire sphere as its support, the proposed model is restricted to the surface of the sphere.}
    \vspace{0pt}
\end{figure}

The frequency-domain sound field $\hat{p}(:=\mathcal{F}_{\mathrm{t}} p)$ can be expressed as an integral over the unit sphere of the wavenumber space (see Appendix\ref{AppendixA}):
\begin{align}
    \hat{p}(\bm{r},\omega) &= \int_{S^{d-1}} \tilde{P}_{\mathrm{f}}\left( \frac{\omega}{c} \bm{\theta} \right) \mathrm{e}^{\mathrm{i} \frac{\omega}{c} \bm{\theta} \cdot \bm{r}} \mathrm{d} \bm{\theta}, \label{eq:herglotz} \\
    \tilde{P}_{\mathrm{f}}\left( \frac{\omega}{c} \bm{\theta} \right) &:= \frac{1}{(2\pi)^{\frac{d-1}{2}}} \left(\frac{\omega}{c}\right)^{d-1} P_{\mathrm{f}}\left( \frac{\omega}{c} \bm{\theta} \right), \label{eq:coefficient_and_spectrum}
\end{align}
where $\bm{\theta}:=(\theta_{1}, \ldots , \theta_{d})$, $\mathrm{d}\bm{\theta}$ is defined by
\begin{equation}
    \mathrm{d}\bm{\theta} := \prod_{i=1}^{d-2} \sin^{d-i-1} \theta_{i} \mathrm{d} \theta_{1} \cdots \mathrm{d} \theta_{d-1}.
\end{equation}
Equations (\ref{eq:herglotz}) and (\ref{eq:coefficient_and_spectrum}) show that the spectral support of the frequency-domain sound field exists only on a hypersphere surface of radius $\omega/c$ and that its spectrum is a constant multiple of the plane forward wave coefficient $P_{\mathrm{f}}(\omega \bm{\theta}/c)$.
The frequency-domain sound field is thus regarded as an element of the space of band-limited functions.
That is, we consider the frequency-domain sound field as an element by constructing a space formed by functions such that the wavenumber spectral support exists only on the hypersphere surface determined by the frequency.
Specifically, we define the function space
\begin{align} \label{eq:def_S_k-d-1}
    &\mathcal{S}_{k}^{d-1} := \left\{ f \in L^{2}(\mathbb{R}^{d}) \mid \exists \tilde{F}: k S^{d-1} \rightarrow \mathbb{C} \right. \notag \\
    &\qquad \qquad \left. \mathrm{s.t.,} \, f(\bm{r}) = \int_{S^{d-1}} \tilde{F} (k \bm{\theta}) \mathrm{e}^{\mathrm{i} k \bm{\theta} \cdot \bm{r}} \mathrm{d} \bm{\theta} \right\},
\end{align}
its inner product is introduced by
\begin{equation}
    \langle f, g \rangle := \int_{S^{d-1}} \tilde{F} (k \bm{\theta}) \overline{\tilde{G} (k \bm{\theta})} \mathrm{d} \bm{\theta},
\end{equation}
and its norm as a natural norm ($\|f\|:=\sqrt{\langle f, f \rangle}$) is defined.
In relation with previous work\cite{Iwami2023}, it means that the wavenumber spectral support changes from a hypersphere to a hypersphere surface (Fig.~\ref{fig1}).
This function space then becomes an RKHS with the RK
\begin{equation} \label{eq:RK}
    \kappa_{k}(\bm{r}, \bm{r}^{\prime}) = 2\pi \left( \frac{2\pi}{k|\bm{r}-\bm{r}^{\prime}|} \right)^{\frac{d}{2}-1} J_{\frac{d}{2}-1}(k |\bm{r}-\bm{r}^{\prime}|),
\end{equation}
where $J_{n}$ denotes the $n$-th order Bessel function.
See Appendix \ref{AppendixB} for the proof.
If the dimensionality of the space is odd, the RK can also be expressed by defining $\tilde{d} := (d-3)/2$:
\begin{equation} \label{eq:RK_of_band-limited_function_space_surface_odd}
    \kappa_{k}(\bm{r}, \bm{r}^{\prime}) = 4\pi \left( \frac{2\pi}{k|\bm{r}-\bm{r}^{\prime}|} \right)^{\tilde{d}} j_{\tilde{d}} (k |\bm{r}-\bm{r}^{\prime}|),
\end{equation}
where $j_{n}$ is the $n$-th order spherical Bessel function.
%--------------------------------------%
%--------------------------------------%
When the field is assumed to have two and three dimensions, the RK is
\begin{equation} \label{eq:2D_k}
    \kappa_{\mathrm{2D}, k}(\bm{r}, \bm{r}^{\prime}) = 2\pi J_{0}(k|\bm{r}-\bm{r}^{\prime}|),
\end{equation}
\begin{equation} \label{eq:3D_k}
    \kappa_{\mathrm{3D}, k}(\bm{r}, \bm{r}^{\prime}) = 4\pi \, j_{0}(k|\bm{r}-\bm{r}^{\prime}|).
\end{equation}
The RK in three-dimensional space in Eq.~(\ref{eq:3D_k}) is equivalent (except for the constant multiplication) to that in the work\cite{Ueno2018} of Ueno \textit{et al}.

Finally, we mention two concerns regarding the discussion thus far.
First, one might think that the function value diverges when $\bm{r} = \bm{r}^{\prime}$ in the expression for the RK (\ref{eq:RK}).
However, returning to the integral form of Eq.~(\ref{eq:kappa_k_integral}), it can be verified that
\begin{equation} \label{eq:RK_zero}
    \kappa_{k}(\bm{r}, \bm{r}) = A_{d-1}.
\end{equation}

Second, so far, the dimensionality $d$ of the space has been assumed to be any natural number.
However, when $d=1$, several equations encounter issues (e.g., the integral over $S^{0}$ in Eqs.~(\ref{eq:herglotz}) and (\ref{eq:def_S_k-d-1})).
The specific issues are the ambiguity of the meaning of the integral over $S^{0}$ that appears in Eq.~( \ref{eq:herglotz}), the consequent need to modify the RKHS, including the inner product, and whether the RK at that time matches that obtained by substituting $d=1$ into Eqs.~( \ref{eq:RK}) and (\ref{eq:RK_of_band-limited_function_space_surface_odd}).
These issues are solved using the following definition of the integral over $S^{0}$.
That is, from $S^{0} = \{ \pm 1 \}$, we define
\begin{equation}
    \int_{S^{0}} f(\theta) \mathrm{d} \theta := \sum_{\theta \in \{ \pm 1 \}} f(\theta).
\end{equation}
Equations~(\ref{eq:herglotz}) and (\ref{eq:def_S_k-d-1}) are then rewritten as
\begin{align}
    \hat{p}(x, \omega) &= \sum_{\theta \in \{ \pm 1 \}} P_{\mathrm{f}} \left( \frac{\omega}{c} \theta \right) \mathrm{e}^{\mathrm{i} \frac{\omega}{c} \theta x}, \\
    \mathcal{S}_{k}^{01} &= \left\{ f \in L^{2}(\mathbb{R}) \mid \exists F: k S^{0} \rightarrow \mathbb{C} \right. \notag \\
    &\qquad \left. \mathrm{s.t.,} \quad f(x) = \sum_{\theta \in \{ \pm 1 \}} F (k \theta) \mathrm{e}^{\mathrm{i} k \theta x} \right\}.
\end{align}
Note that $\tilde{P}_{\mathrm{f}} = P_{\mathrm{f}}$ holds when $d=1$.
Accordingly, the inner product of $f, g \in \mathcal{S}_{k}^{0}$ is rewritten as
\begin{equation}
\langle f, g \rangle = \sum_{\theta \in { \pm 1 }} F (k \theta) \overline{G (k \theta)}.
\end{equation}
The RK can be calculated as
\begin{equation}
    \kappa_{k}(x, x^{\prime}) = \sum_{\theta \in \{ \pm 1 \}} \mathrm{e}^{\mathrm{i} k \theta (x - x^{\prime})} = 2 \cos k (x - x^{\prime}).
\end{equation}
This equation needs to be derived similarly from Eq.~(\ref{eq:RK}) and Eq.~(\ref{eq:RK_of_band-limited_function_space_surface_odd}), which can be immediately verified from
\begin{align}
    J_{-\frac{1}{2}}(z) &= \sqrt{\frac{2}{\pi z}} \cos z, \\
    j_{-1}(z) &= - \frac{\cos z}{z}.
\end{align}

%--------------------------------------%
\section{\label{sec:3}Sound field expression and wavenumber spectrum estimation}
%--------------------------------------%
This section describes the reconstruction of the frequency-domain sound field $\hat{p}(\cdot,\omega) \in \mathcal{S}_{k}^{d-1}$ and estimation of its wavenumber spectrum $\hat{P}(\cdot,\omega)(:=\mathcal{F}_{\mathrm{s}}\hat{p}(\cdot,\omega))$.
For simplicity, the angular frequency index $\omega$ may be omitted hereafter.

\begin{figure}[t]
    \centering
    \includegraphics[width=0.55\columnwidth]{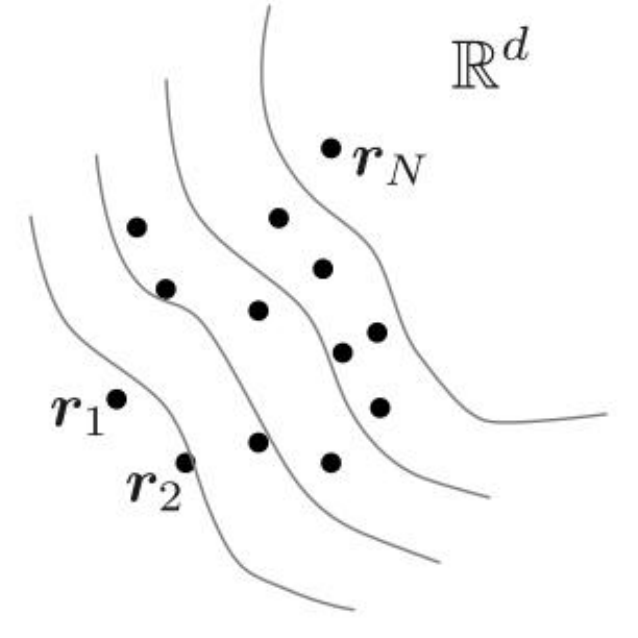} \vspace{0pt}
    \caption{\label{fig2}Problem setting in Sec.~\ref{sec:3}.}
    \vspace{0pt}
\end{figure}

Under the assumption that the sound field is free, we consider the sampling problem with sampling points $\{ \bm{r}_{n} \}_{n=1}^{N} \subset \mathbb{R}^{d}$ (Fig.~\ref{fig2}).
The sampled values $\{ \hat{p}_{n} \}_{n=1}^{N} \subset \mathbb{C}$ are given by
\begin{equation}
    \hat{p}_{n} = \hat{p}(\bm{r}_{n}) + \varepsilon_{n} \quad (n=1, 2, \ldots, N),
\end{equation}
where $\varepsilon_{n}$ is random noise with zero mean.

%--------------------------------------%
%--------------------------------------%

%------------------
\subsection{\label{sec:3.1}Sound field reconstruction}
Let $A$ be the sampling operator; i.e.,
\begin{equation}
    A\hat{p} := \hat{\bm{p}},
\end{equation}
where $\hat{\bm{p}} := [\hat{p}(\bm{r}_{1}) \cdots \hat{p}(\bm{r}_{N})]^{\mathsf{T}} \in \mathbb{C}^{N}$.
If an unbiased reconstruction operator $X:\mathbb{C}^{N} \rightarrow \mathcal{S}_{k}^{d-1}$ can be constructed such that $XA=I$, then a complete reconstruction can be achieved on average; i.e., $XA\hat{p} = \hat{p}$.
However, the reconstruction is unachievable because the dimensionality of $\mathcal{S}_{k}^{d-1}$ is infinite and the number of sampling points are finite.

In this case, we should find the largest orthogonally projectable subspace $S$ of $\mathcal{S}_{k}^{d-1}$. It was shown in \cite{Ogawa2007} that such $S$ is $\mathcal{R}(A^{\ast})$ (where $\mathcal{R}(A^{\ast})$ denotes the range of the adjoint operator of $A$).
The fact that such an unbiased operator $X$ (i.e., $XA = P_{\mathcal{R}(A^{\ast})}$) satisfies
\begin{equation} \label{eq:reconstruction_op}
    X \hat{\bm{p}} = \bm{\kappa}^{\mathsf{T}} \mathcal{K}^{-1} \hat{\bm{p}}
\end{equation}
is shown in \cite{Shimamura2009}.
Here, $\bm{\kappa} := [\kappa_{k}(\cdot, \bm{r}_{1}) \cdots \kappa_{k}(\cdot, \bm{r}_{N})]^{\mathsf{T}}$ and $\mathcal{K}$ denotes the Gram matrix $\mathcal{K} = (\kappa_{k}(\bm{r}_{i}, \bm{r}_{j}))$.
Equation~(\ref{eq:reconstruction_op}) implies that the best approximation of $\hat{p}$ in the noiseless situation has the form
\begin{equation} \label{eq:RK_model}
    \hat{p}_{\mathrm{est}} = \sum_{n=1}^{N} a_{n} \kappa_{k}(\cdot, \bm{r}_{n}),
\end{equation}
and its coefficient vector $\bm{a} := [a_{1} \cdots a_{N}]^{\mathsf{T}}$ is then 
\begin{equation} \label{eq:coefficient_noiseless}
    \bm{a} = \mathcal{K}^{-1} \hat{\bm{p}}.
\end{equation}
Under noisy conditions, regularization is effective, and the coefficient vector when using Tikhonov regularization\cite{Shimamura2009, Hansen2010} is 
\begin{equation} \label{eq:coefficient_Tikhonov}
    \bm{a} = (\mathcal{K} + \lambda I)^{-1} \hat{\bm{p}}.
\end{equation}
Therefore, the estimation of the sound pressure at arbitrary positions $\{ \bm{r}_{m}^{\prime} \}_{m=1}^{M} \subset \mathbb{R}^{d}$ is realized as 
\begin{equation} \label{eq:interpolation}
    \hat{\bm{p}}_{\mathrm{est}} = \mathcal{K}^{\prime} (\mathcal{K} + \lambda I_{N})^{-1} \hat{\bm{p}},
\end{equation}
where
\begin{align}
    \hat{\bm{p}}_{\mathrm{est}} &:= \left[ \begin{array}{ccc}
    \hat{p}_{\mathrm{est}}(\bm{r}_{1}^{\prime}) & \cdots & \hat{p}_{\mathrm{est}}(\bm{r}_{M}^{\prime})
    \end{array} \right]^{\mathsf{T}}, \\
    \mathcal{K}^{\prime} = (\mathcal{K}^{\prime}_{i,j}) &:= (\kappa_{\frac{\omega}{c}}(\bm{r}_{i}^{\prime}, \bm{r}_{j})).
\end{align}
It is seen from the above discussion that at any sampling points, the best reconstruction of the original function is in the form of Eq.~(\ref{eq:RK_model}); i.e., the RK is taken as the system of functions.

Finally, it is worth noting kernel ridge regression (KRR)\cite{Shawe2010}.
KRR on $\mathcal{S}_{k}^{d-1}$ is the optimization problem
\begin{equation} \label{eq:KRR}
    \min_{f \in \mathcal{S}_{k}^{d-1}} \sum_{n=1}^{N} | p_{n} - f(\bm{r}_{n}) |^{2} + \lambda \| f \|^{2}.
\end{equation}
Its solution is Eqs.~(\ref{eq:RK_model}) and (\ref{eq:coefficient_Tikhonov}).
That is, KRR gives the same solution.
However, KRR minimizes only the error of the function values at the sampling points and does not guarantee proximity as a function.

%--------------------------------------%
\begin{table*}[t]
    \caption{\label{table1}Differences between the existing model \cite{Iwami2023} and proposed model}
    \centering
    \begin{tabular}{|@{\hspace{3mm}}l@{\hspace{3mm}}||@{\hspace{3mm}}c@{\hspace{3mm}}|@{\hspace{3mm}}c@{\hspace{3mm}}|@{\hspace{3mm}}c@{\hspace{3mm}}|@{\hspace{3mm}}c@{\hspace{3mm}}|}
        \hline
        & Spectral support & Reproducing kernel & Wavenumber spectrum & Sampling \\
        \hline
        Previous model & $B_{K}(\bm{0})$ & $\left(\frac{K}{2\pi |\bm{r}-\bm{r}^{\prime}|}\right)^{\frac{d}{2}} J_{\frac{d}{2}} (K|\bm{r}-\bm{r}^{\prime}|)$ & Axis & Dense \\
        \hline
        Proposed model & $kS^{d-1}$ & $2\pi \left( \frac{2\pi}{k|\bm{r}-\bm{r}^{\prime}|} \right)^{\frac{d}{2}-1} J_{\frac{d}{2}-1}(k |\bm{r}-\bm{r}^{\prime}|)$ & Direction & Sparse \\        \hline
    \end{tabular}
\end{table*}
%--------------------------------------%

%------------------
\subsection{\label{sec:3.2}Estimation of the wavenumber spectrum}
The estimation of the wavenumber spectrum is simplified by considering the sound field as an element of the band-limited functions.
As the sound field is now estimated in the form of Eq.~(\ref{eq:RK_model}), it can be transformed by applying a spatial Fourier transform:
\begin{equation} \label{eq:spectrum_estimation_def}
    \hat{P}_{\mathrm{est}} = \sum_{n=1}^{N} a_{n} \mathcal{F}_{\mathrm{s}} \kappa_{k}(\cdot, \bm{r}_{n}).
\end{equation}
The RK can be rewritten in integral form as
\begin{align}
    &\kappa_{k}(\bm{r}, \bm{r}^{\prime}) = \int_{S^{d-1}} \mathrm{e}^{\mathrm{i} k \bm{\theta}\cdot (\bm{r} - \bm{r}^{\prime})} \mathrm{d}\bm{\theta} \notag \\
    &\mathalpha{=} \int_{S^{d\mathalpha{-}1}} \!\! \int_{\mathbb{R}_{+}} k^{\prime 1\mathalpha{-}d} \mathrm{e}^{\mathrm{i} k^{\prime} \bm{\theta}\cdot (\bm{r} \mathalpha{-} \bm{r}^{\prime})} \delta(k \mathalpha{-} k^{\prime}) k^{\prime d\mathalpha{-}1} \mathrm{d}k^{\prime} \mathrm{d}\bm{\theta} \notag \\
    &\mathalpha{=} \frac{1}{(2\pi)^{\frac{d}{2}}} \int_{\mathbb{R}^{d}} (2\pi)^{\frac{d}{2}} k^{\prime 1\mathalpha{-}d} \mathrm{e}^{\mathalpha{-}\mathrm{i} \bm{k}^{\prime} \cdot \bm{r}^{\prime}} \delta(k \mathalpha{-} k^{\prime}) \mathrm{e}^{\mathrm{i} \bm{k}^{\prime} \cdot \bm{r}} \mathrm{d} \bm{k}^{\prime}.
\end{align}
Applying the spatial Fourier transform, Eq.~(\ref{eq:spectrum_estimation_def}) is reformulated as
\begin{equation} \label{eq:spectrum_estimation}
    \hat{P}_{\mathrm{est}}\left( k \bm{\theta} \right) = (2\pi)^{\frac{d}{2}} k^{1-d} \delta(k - k^{\prime}) \sum_{n=1}^{N} a_{n} \mathrm{e}^{-\mathrm{i} k \bm{\theta} \cdot \bm{r}_{n}}.
\end{equation}
This equation can be used to estimate the wavenumber spectrum of the frequency-domain sound field.
By comparing with Eq.~(\ref{eq:rel_P_hat_and_P_f}), the plane forward wave coefficient $P_{\mathrm{f}}$ is estimated according to
\begin{equation} \label{eq:Pf_estimation_freq}
    P_{\mathrm{f,est}}(k\bm{\theta}) = (2\pi)^{\frac{d-1}{2}} k^{1-d} \sum_{n=1}^{N} a_{n}  \mathrm{e}^{-\mathrm{i} \bm{k} \cdot \bm{r}_{n}}.
\end{equation}
As a computer cannot handle the Dirac delta function, only the forward wave coefficient can be estimated.
However, this coefficient can be considered as essentially the wavenumber spectrum.

If we want to estimate for any directions $\{ \bm{\theta}_{m}^{\prime} \}_{m=1}^{M} \subset \mathbb{S}^{d-1}$, we use the matrix operation
\begin{equation} \label{eq:wavenumber_spectrum_estimation}
    \bm{P}_{\mathrm{f,est}} = \mathcal{L} (\mathcal{K} + \lambda I_{N})^{-1} \hat{\bm{p}},
\end{equation}
where
\begin{align}
    \bm{P}_{\mathrm{f,est}} &:= \left[ \begin{array}{ccc}
    P_{\mathrm{f,est}}\left( k \bm{\theta}_{1}^{\prime} \right) & \cdots & P_{\mathrm{f,est}}\left( k \bm{\theta}_{M}^{\prime} \right)
    \end{array} \right]^{\mathsf{T}}, \\
    \mathcal{L} = (\mathcal{L}_{i,j}) &:= \left( (2\pi)^{\frac{d-1}{2}} k^{1-d} \mathrm{e}^{-\mathrm{i} k \bm{\theta}_{i}^{\prime} \cdot \bm{r}_{j}} \right).
\end{align}

%------------------
\subsection{\label{sec:3.3}Comparison of models between the instantaneous sound field and frequency-domain sound field}
As mentioned earlier, the proposed modeling follows a framework similar to that used in previous work \cite{Iwami2023}.
In this section, the differences between the proposed modeling and previous modeling are discussed.

The comparison is summarized in Table~\ref{table1}.
The support of the wavenumber spectrum is restricted to a hypersphere of radius $K$ for instantaneous sound fields with an upper wavenumber limit $K$, and to a hyperspherical surface of radius $k$ for wavenumbers $k$ in the frequency-domain sound fields (Fig.~\ref{fig1}).
This can be interpreted as the dimensionality of the manifold of support being reduced by one.
As a result, the RK is constructed using Bessel functions of one order lower.

For the wavenumber spectrum of the sound field, in the previous model, Eq.~(\ref{eq:wavenumber_spectrum_t}) is estimated using only the spatial information at time $t$.
As this equation involves a mixture of coefficients for forward and backward waves, information on the direction of sound arrival cannot be obtained, and only the axis of propagation can be obtained.
Meanwhile, in the proposed model, as the wavenumber spectrum and plane forward wave coefficient have almost the same meaning, information on the direction of arrival can be obtained.

Let us consider the sampling points (microphone placements).
The previous model is based on the classical sampling theorem in $d$-dimensional space.
Hence, the adoption of lattice sampling, for example, guarantees good modeling performance.

Meanwhile, the optimal placement problem for the proposed model is still not well understood.
However, for the spherical harmonic expansion model of sound fields, it is well known that spherical arrangements are effective \cite{Boaz2015}, and several specific placement methods have been proposed \cite{Fliege1999, Lecomte2016}.
The proposed model provides the best approximation regardless of the arrangement, so it should also yield good approximations for such effective arrangements.
In this sense, the frequency-domain sound field model provides a good approximation with sampling that is spatially sparse compared with that for the instantaneous field model.

%--------------------------------------
\section{\label{sec:4}Numerical simulation}
%--------------------------------------
%--------------------------------------%
\begin{table}[t]
    \caption{\label{table2}Conditions in numerical simulations}
    \centering
    \begin{tabular}{l@{\hspace{15mm}}l}
        \hline\hline
        Frequency & 2000~Hz \\
        Direction of travel & 45$^{\circ}$ \\
        Side length of square area & 0.4~m\\
        Sampling points $N$ & 21~samples \\
        Signal-to-noise ratio & 30~dB\\
        Regularization parameter $\lambda$ & 0.01 \\
        \hline\hline
    \end{tabular}
\end{table}
%--------------------------------------%
Two simple numerical simulations of a two-dimensional sound field were conducted.
One simulation relates to estimation of a sound field and the other to estimation of a wavenumber spectrum.

The common parameters of the two simulations are given in Table~\ref{table2}.
We used a plane wave sound with a frequency of 2000~Hz propagating in the direction of 45$^{\circ}$ as the sound field.
The 21 sampling points were distributed randomly using a uniform distribution over a square with a side length of 0.4~m.
For practicality, the signal-to-noise ratio was set at 30~dB, and Tikhonov regularization with a regularization parameter of 0.01 was adopted for the inverse problems.

%------------------
\subsection{\label{sec:4.1}Comparison with the existing method}
To verify the performance of the proposed model in representing the sound field, we conducted a simulation comparing the sound field estimation performance of the proposed model with that of a well-established model, namely the spherical harmonic expansion model.
As we now assume a two-dimensional sound field, the spherical harmonic expansion model for the sound field $\hat{p}$ up to order $N$ is given by
\begin{equation}
    \hat{p}(r, \theta) \approx \sum_{n=-N}^{N} b_{n} J_{|n|}(kr) \mathrm{e}^{\mathrm{i}n\theta},
\end{equation}
where $b_{n} \, (n=-N, \ldots, N)$ denotes the expansion coefficient.
As there are 21 sampling points, the maximum order $N$ is determined to be 10.
The coefficients are determined by simultaneously solving the equations for all sampling points adopting Tikhonov regularization.

The normalized error (NE) (i.e., difference) between the reference sound field $\hat{p}(\bm{r})$ and the estimated sound field $\hat{p}_{\mathrm{est}}(\bm{r})$ as an index is defined by
\begin{equation} \label{eq:NE}
	\mathrm{NE}(\bm{r}) = 20 \log_{10} \frac{|\hat{p}(\bm{r})-\hat{p}_{\mathrm{est}}(\bm{r})|}{|\hat{p}(\bm{r})|}.
\end{equation}

\begin{figure*}[t]
    \centering
    \includegraphics[width=1.8\columnwidth]{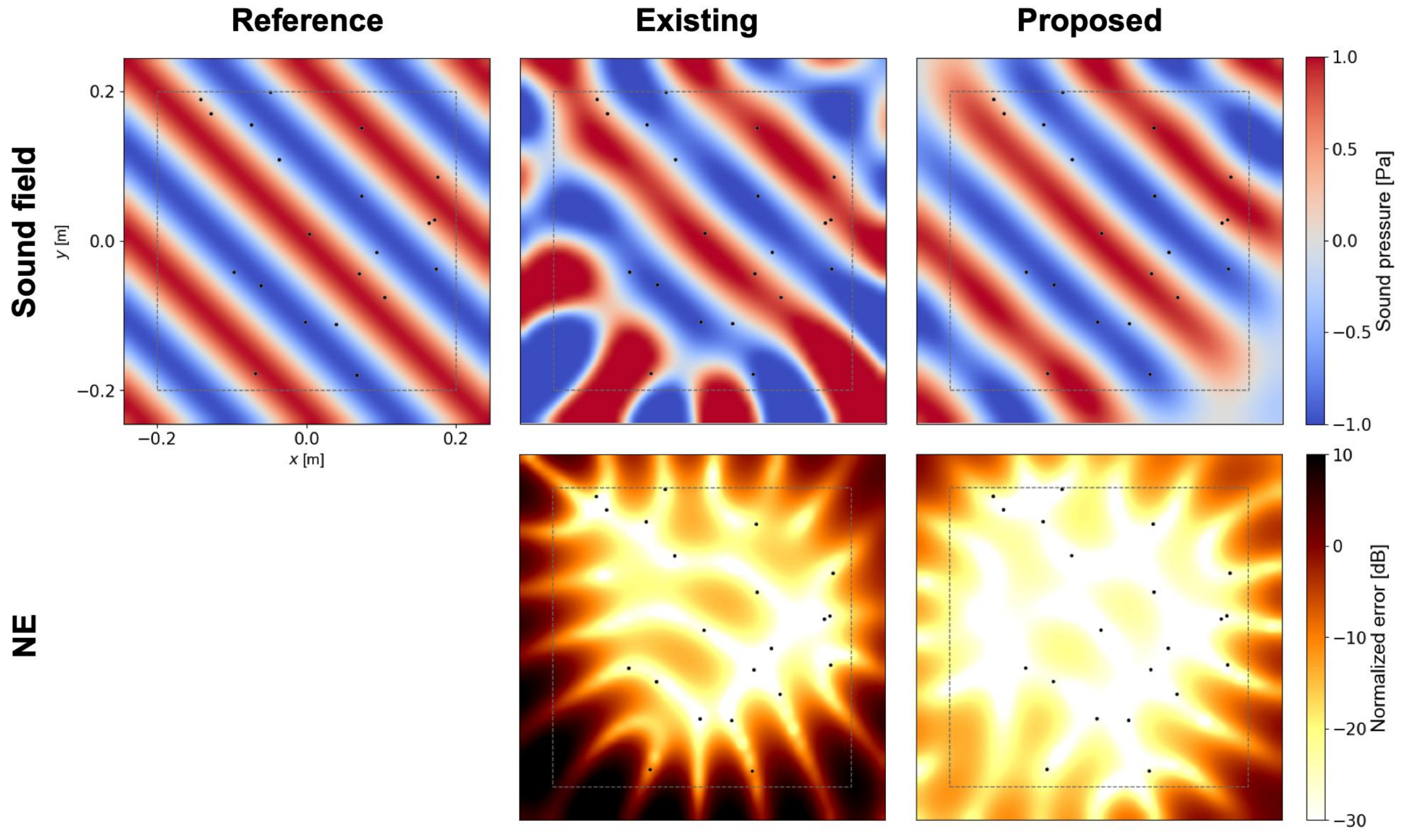} \vspace{0pt}
    \caption{\label{fig3}Estimated sound fields and their NE distributions when using the existing and proposed models.}
    \vspace{0pt}
\end{figure*}

The estimates obtained using the existing model and proposed model are shown in Fig.~\ref{fig3}.
The black dots indicate sampling positions and the region enclosed by dashed squares represents the area where there are potentially sampling points.
The average NEs within this area for the existing model and proposed model are $-16.0$ and $-25.3$~dB, respectively.
In other words, on average, the proposed method is more accurate than the existing method in estimating the sound field.
In particular, it is seen that the performance of the existing method is poor in regions with few sampling points.
These results imply that the proposed model has a high approximation capability.

%--------------------------------------%
%--------------------------------------%

%------------------
\subsection{\label{sec:4.2}Wavenumber spectrum estimation}
Under the same conditions as in the previous simulation, we conducted an estimation of the wavenumber spectrum using Eq.~(\ref{eq:wavenumber_spectrum_estimation}).

\begin{figure}[t]
    \centering
    \includegraphics[width=1.0\columnwidth]{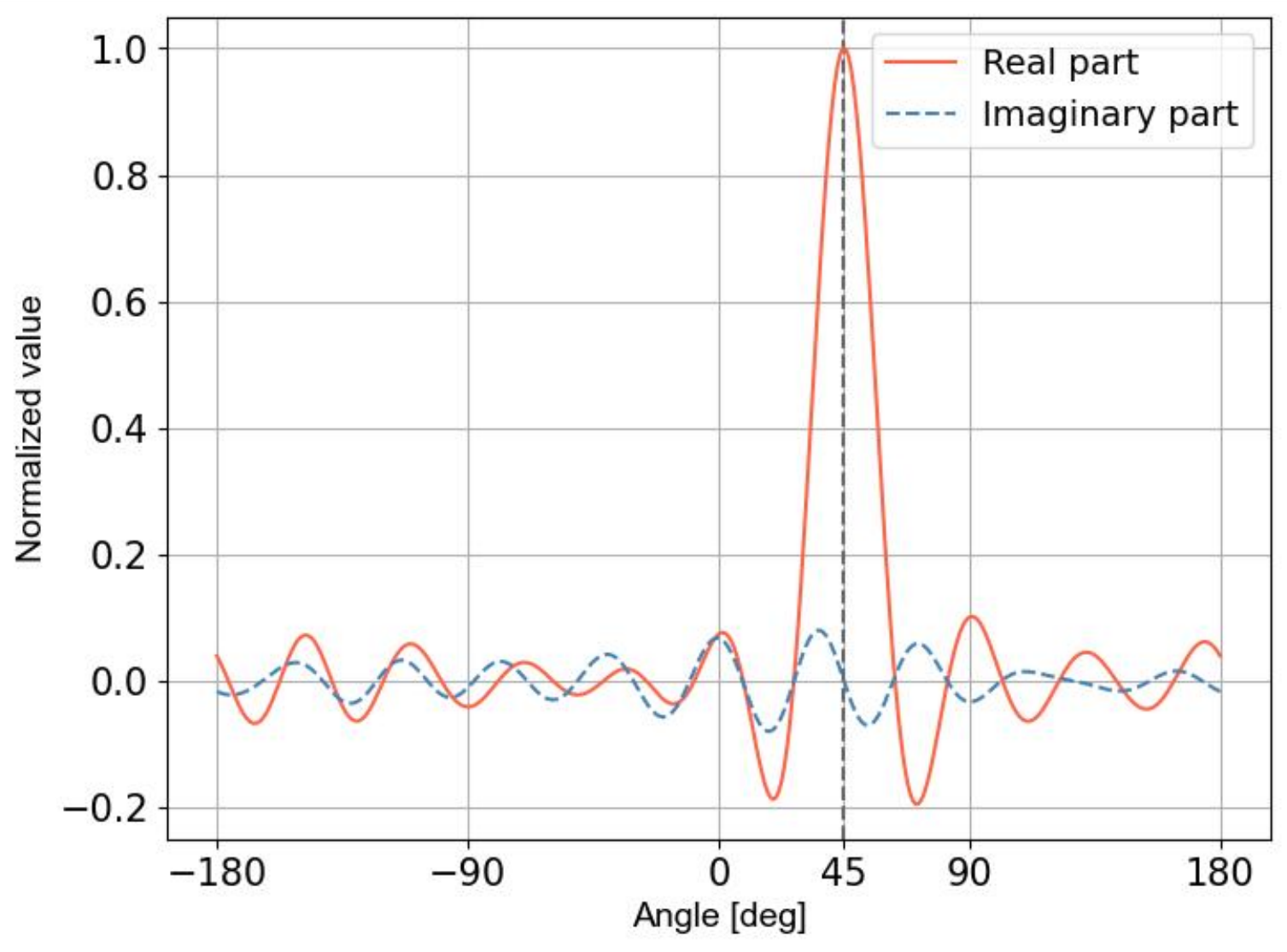} \vspace{0pt}
    \caption{\label{fig4}Result of the proposed estimation of the wavenumber spectrum.}
    \vspace{0pt}
\end{figure}

The estimated results are illustrated in Fig.~\ref{fig4}.
The sound field is now a plane wave traveling in the direction of $45^{\circ}$.
Therefore, the real part should be non-zero exclusively at $45^{\circ}$ and should be zero at other angles (strictly speaking, it should resemble a Dirac delta function).
Simultaneously, the imaginary part should be zero irrespective of the angle.
The estimation results exhibit a peak in the real part at $45^{\circ}$, whereas the real part at other angles and the imaginary part are close to zero.
In this sense, the overall shape is captured reasonably well.
This result suggests that the estimation of the wave number spectrum based on the best approximation model of the sound field is effective.
Furthermore, applications of the proposed method such as the estimation of the direction of arrival are conceivable.

A well-known method for estimating the wave number spectrum is spatial discrete Fourier transformation adopting grid sampling.
Whereas this established method requires a dense array of sampling points (i.e., a dense array of microphones), the proposed method has the advantage of adopting arbitrary sampling.

%--------------------------------------%
%--------------------------------------%

%--------------------------------------
\section{\label{sec:5}Concluding remarks}
%--------------------------------------
This paper proposed a representation of a frequency-domain sound field in general dimensions by considering the field as an element of a band-limited function space.
Specifically, we leveraged the fact that the support of the wavenumber spectrum of the frequency-domain sound field is restricted to the surface of a sphere.
We constructed such a band-limited space and represented the sound field as a linear combination using its RK.
We also discussed how this model performs well in approximating the sound field in the problem of sound field reconstruction.
The proposed model aligns with the work of Ueno \textit{et al.} for three dimensions; thus, our work can be recognized as a generalization in terms of dimensionality.
One advantage of our model lies in the ease of formulating the estimation of the wavenumber spectrum, which is attributed to the spatial inverse Fourier transform representation of the RK.

Two numerical experiments were conducted using a simple two-dimensional sound field.
One aimed to verify the sound field estimation performance, adopting the spherical harmonic expansion model of the sound field for comparison.
The proposed model gave generally better estimation results than the existing model, implying its practical effectiveness.
The second was a simulation conducted to verify the validity of the wavenumber spectrum estimation.
Results obtained using the proposed method showed a strong peak in the direction of plane-wave propagation.

Future works will include experiments using actual microphone arrays and the application of the proposed model to numerical simulation.

\begin{acknowledgments}
This research was partially supported by JSPS KAKENHI under Grant Numbers JP21H03764 and JP24K03222. We thank Edanz (https://jp.edanz.com/ac) for editing a draft of this manuscript.
\end{acknowledgments}

% -------------------------------------------------------------------------------------------------------------------
%   Appendix  (optional)
\appendix
%--------------------------------------
\section{\label{AppendixA} Proof that the wavenumber spectral support of the sound field is restricted on a unit sphere}
%--------------------------------------
The conclusion of equation (\ref{eq:herglotz}) is verified as follows:
\begin{align}
    &\hat{P}(\bm{k}, \omega) = \mathcal{F}_{\mathrm{t}} \mathcal{F}_{\mathrm{s}} p(\bm{k}, \omega) \notag \\
    &= \frac{1}{\sqrt{2\pi}} \int_{\mathbb{R}} \left( \overline{P_{\mathrm{b}}(\bm{k})} \mathrm{e}^{\mathrm{i}c|\bm{k}|t} + P_{\mathrm{f}}(\bm{k}) \mathrm{e}^{- \mathrm{i}c|\bm{k}|t} \right) \mathrm{e}^{\mathrm{i} \omega t}  \notag \\
    &= \frac{1}{\sqrt{2\pi}}\overline{P_{\mathrm{b}}(\bm{k})} \int_{\mathbb{R}} \mathrm{e}^{\mathrm{i} (\omega + c|\bm{k}|)t} \mathrm{d} t \notag \\
    &\qquad \quad + \frac{1}{\sqrt{2\pi}}P_{\mathrm{f}}(\bm{k}) \int_{\mathbb{R}} \mathrm{e}^{\mathrm{i} (\omega - c|\bm{k}|)t} \mathrm{d} t \notag \\
    &= \sqrt{2\pi} \left\{ \overline{P_{\mathrm{b}}(\bm{k})} \delta(\omega + c|\bm{k}|) + P_{\mathrm{f}}(\bm{k}) \delta(\omega - c|\bm{k}|) \right\} \notag. 
\end{align}
As the first term on the right-hand side can be ignored from $\omega \in \mathbb{R}_{+}$,
\begin{equation} \label{eq:rel_P_hat_and_P_f}
    \hat{P}(\bm{k}, \omega) = \sqrt{2\pi} P_{\mathrm{f}}(\bm{k}) \delta(\omega - c|\bm{k}|).
\end{equation}
The sound field can then be rewritten as
\begin{align}
    &\hat{p}(\bm{r}, \omega) = \mathcal{F}_{\mathrm{s}}^{-1} \hat{P} (\bm{r}, \omega) \notag \\
    &= \frac{1}{(2\pi)^{\frac{d}{2}}} \int_{\mathbb{R}^{d}} \hat{P}(\bm{k}, \omega) \mathrm{e}^{\mathrm{i} \bm{k} \cdot \bm{r}} \mathrm{d} \bm{k} \notag \\
    &= \frac{1}{(2\pi)^{\frac{d-1}{2}}} \int_{\mathbb{R}^{d}} P_{\mathrm{f}}(\bm{k}) \delta(\omega - c|\bm{k}|) \mathrm{e}^{\mathrm{i} \bm{k} \cdot \bm{r}} \mathrm{d} \bm{k} \notag \\
    &= \frac{1}{(2\pi)^{\frac{d-1}{2}}} \notag \\
    &\times \int_{S^{d-1}} \int_{\mathbb{R}_{+}} P_{\mathrm{f}}(k\bm{\theta}) \delta(\omega - ck) \mathrm{e}^{\mathrm{i} k \bm{\theta} \cdot \bm{r}} k^{d-1} \mathrm{d} k \mathrm{d} \bm{\theta} \notag \\
    &= \frac{1}{(2\pi)^{\frac{d-1}{2}}} \left(\frac{\omega}{c}\right)^{d-1} \int_{S^{d-1}} P_{\mathrm{f}}\left( \frac{\omega}{c} \bm{\theta} \right) \mathrm{e}^{\mathrm{i} \frac{\omega}{c} \bm{\theta}\cdot\bm{r}} \mathrm{d}\bm{\theta} \notag \\
    &= \int_{S^{d-1}} \tilde{P}_{\mathrm{f}}\left( \frac{\omega}{c} \bm{\theta} \right) \mathrm{e}^{\mathrm{i} \frac{\omega}{c} \bm{\theta} \cdot \bm{r}} \mathrm{d} \bm{\theta}. \notag
\end{align}

%--------------------------------------
\section{\label{AppendixB} Derivation of the RK}
%--------------------------------------
The RK $\kappa_{k}$ is defined as a map satisfying $\kappa_{k}(\cdot, \bm{r}^{\prime}) \in \mathcal{S}_{k}^{d-1} \quad (\forall \bm{r}^{\prime} \in \mathbb{R}^{d})$ and $\forall f \in \mathcal{S}_{k}^{d-1} \Longrightarrow \langle f, \kappa_{k}(\cdot, \bm{r}^{\prime}) \rangle \mathalpha{=} f(\bm{r}^{\prime}) \quad (\forall \bm{r}^{\prime} \in \mathbb{R}^{d})$.

If $f \in \mathcal{S}_{k}^{d-1}$, then from the definition of space,
\begin{equation} \label{eq:f}
    f(\bm{r}^{\prime}) = \int_{S^{d-1}} \tilde{F} (k \bm{\theta}) \mathrm{e}^{\mathrm{i} k \bm{\theta} \cdot \bm{r}^{\prime}} \mathrm{d} \bm{\theta}
\end{equation}
holds.
$\kappa_{k}$ is now defined as
\begin{equation} \label{eq:kappa_k_integral}
    \kappa_{k}(\bm{r}, \bm{r}^{\prime}) := \int_{S^{d-1}} \mathrm{e}^{\mathrm{i} k \bm{\theta} \cdot (\bm{r} - \bm{r}^{\prime})} \mathrm{d} \bm{\theta}.
\end{equation}
The right-hand side of Eq.~(\ref{eq:f}) can be rewritten as $\langle f, \kappa_{k}(\cdot, \bm{r}^{\prime}) \rangle$.
That is, the RK is given by Eq.~(\ref{eq:kappa_k_integral}).
The specific form of the RK is verified by
\begin{align}
    &\int_{S^{d-1}} \mathrm{e}^{\mathrm{i} k \bm{\theta}\cdot\bm{r}} \mathrm{d}\bm{\theta} = \int_{S^{d-1}} \cos \left( k \bm{\theta}\cdot\bm{r} \right) \mathrm{d}\bm{\theta} \notag \\
    &= \int_{0}^{2\pi} \int_{0}^{\pi} \cdots \int_{0}^{\pi}  \cos \left( kr \cos \theta_{1} \right) \notag \\
    &\qquad \times \prod_{i=1}^{d-2} \sin^{d-i-1} \theta_{i} \mathrm{d} \theta_{1} \cdots \mathrm{d} \theta_{d-1} \notag \\
    &= \int_{0}^{2\pi} \int_{0}^{\pi} \cdots \int_{0}^{\pi} \prod_{i=2}^{d-2} \sin^{d-i-1} \theta_{i} \mathrm{d} \theta_{2} \cdots \mathrm{d} \theta_{d-1} \notag \\
    &\qquad \times \int_{0}^{\pi} \cos \left( k r \cos \theta \right) \sin^{d-2} \theta \mathrm{d}\theta \notag \\
    &= A_{d-2} \int_{0}^{\pi} \cos \left( k r \cos \theta \right) \sin^{d-2} \theta \mathrm{d}\theta \notag \\
    &= \frac{2 \pi^{\frac{d-1}{2}}}{\Gamma(\frac{d-1}{2})} \int_{0}^{\pi} \cos \left( kr \cos \theta \right) \sin^{d-2} \theta \mathrm{d}\theta, \notag
\end{align}
where $A_{n-1}:=2\pi^{n/2}/\Gamma(n/2)$ denotes the measure of the $n-1$-dimensional unit ball surface (where $\Gamma$ is the Gamma function).
As $d^{\prime} := d/2-1$, we have
\begin{align}
    &= \frac{ 2\pi^{d^{\prime} + \frac{1}{2}}}{\Gamma(d^{\prime} + \frac{1}{2})} \int_{0}^{\pi} \cos \left( kr \cos \theta \right) \sin^{2d^{\prime}} \theta \mathrm{d}\theta \notag \\
    &= 2\pi \left( \frac{2\pi}{kr} \right)^{d^{\prime}} \notag \\
    &\times \frac{\left( \frac{kr}{2} \right)^{d^{\prime}}}{\sqrt{\pi} \Gamma(d^{\prime} + \frac{1}{2})} \int_{0}^{\pi} \cos \left( k r \cos \theta \right) \sin^{2d^{\prime}} \theta \mathrm{d}\theta \notag \\
    &= 2\pi \left( \frac{2\pi}{kr} \right)^{d^{\prime}} J_{d^{\prime}}(kr) \notag \\
    &= 2\pi \left( \frac{2\pi}{kr} \right)^{\frac{d}{2}-1} J_{\frac{d}{2}-1}(kr). \notag
\end{align}
The Poisson integral representation of the Bessel function\cite{Nikiforov1988} was used in the equation transformation.

%=======================================================

%=======================================================


\begin{thebibliography}{10}
    \def\enquote#1,{``#1,''}
    \def\enxquote#1{``#1''}
    \expandafter\ifx\csname url\endcsname\relax
      \def\url#1{\texttt{#1}}\fi
    \expandafter\ifx\csname urlprefix\endcsname\relax\def\urlprefix{URL }\fi
    \providecommand{\bibinfo}[2]{#2}
    \def\plainquote#1{``#1''}
    \providecommand{\noopsort}[1]{}
    \providecommand{\switchargs}[2]{#2#1}
    \providecommand{\dourl}[1]{\href{http://#1}{\nolinkurl{#1}}}
      \def\eatspace #1{#1}
    
    \bibitem{Berkhout1993}
    \bibinfo{author}{A.~J. Berkhout}, \bibinfo{author}{D.~de~Vries}, and \bibinfo{author}{P.~Vogel}, \enquote{\bibinfo{title}{Acoustic control by wave field synthesis}},  \bibinfo{journal}{J. Acoust. Soc. Am.} \textbf{93}(5), \bibinfo{pages}{2764--2778} (\bibinfo{year}{1993}).
    
    \bibitem{Spors2008}
    \bibinfo{author}{S.~Spors}, \bibinfo{author}{R.~Rabenstein}, and \bibinfo{author}{J.~Ahrens}, \enquote{\bibinfo{title}{The theory of wave field synthesis revisited}}, in \emph{\bibinfo{booktitle}{Proc. 124th AES Conv.}} (\bibinfo{year}{2008}), pp. \bibinfo{pages}{17--20}.
    
    \bibitem{Cooper1972}
    \bibinfo{author}{D.~H. Cooper} and \bibinfo{author}{T.~Shiga}, \enquote{\bibinfo{title}{Discrete-matrix multichannel stereo}},  \bibinfo{journal}{J. Audio Eng. Soc.} \textbf{20}(5), \bibinfo{pages}{346--360} (\bibinfo{year}{1972}).
    
    \bibitem{Gerzon1973}
    \bibinfo{author}{M.~A. Gerzon}, \enquote{\bibinfo{title}{Periphony: With-height sound reproduction}},  \bibinfo{journal}{J. Audio Eng. Soc.} \textbf{21}(1), \bibinfo{pages}{2--10} (\bibinfo{year}{1973}).
    
    \bibitem{Poletti2005}
    \bibinfo{author}{M.~A. Poletti}, \enquote{\bibinfo{title}{Three-dimensional surround sound systems based on spherical harmonics}},  \bibinfo{journal}{J. Audio Eng. Soc.} \textbf{53}(11), \bibinfo{pages}{1004--1025} (\bibinfo{year}{2005}).
    
    \bibitem{Ihlenburg1998}
    \bibinfo{author}{F.~Ihlenburg}, \emph{\bibinfo{title}{Finite element analysis of acoustic scattering}}  (\bibinfo{publisher}{Springer}, \bibinfo{address}{New York}, \bibinfo{year}{1998}).
    
    \bibitem{Marburg2008}
    \bibinfo{author}{S.~Marburg} and \bibinfo{author}{B.~Nolte}, \emph{\bibinfo{title}{Computational acoustics of noise propagation in fluids: finite and boundary element methods}}  (\bibinfo{publisher}{Springer-Verlag}, \bibinfo{address}{Berlin}, \bibinfo{year}{2008}).
    
    \bibitem{Bishop2006}
    \bibinfo{author}{C.~M. Bishop}, \emph{\bibinfo{title}{Pattern recognition and machine learning}}  (\bibinfo{publisher}{Springer}, \bibinfo{address}{New York}, \bibinfo{year}{2006}).
    
    \bibitem{Ueno2018}
    \bibinfo{author}{N.~Ueno}, \bibinfo{author}{S.~Koyama}, and \bibinfo{author}{H.~Saruwatari}, \enquote{\bibinfo{title}{Kernel ridge regression with constraint of helmholtz equation for sound field interpolation}}, in \emph{\bibinfo{booktitle}{2018 16th International Workshop on Acoustic Signal Enhancement (IWAENC)}} (\bibinfo{year}{2018}), pp. \bibinfo{pages}{1--440}.
    
    \bibitem{Iwami2023}
    \bibinfo{author}{T.~Iwami}, \bibinfo{author}{K.-i. Sawai}, and \bibinfo{author}{A.~Omoto}, \enquote{\bibinfo{title}{Direction-of-arrival estimation in half-space from single sample array snapshot}},  \bibinfo{journal}{J. Acoust. Soc. Am.} \textbf{153}(5), \bibinfo{pages}{3025--3025} (\bibinfo{year}{2023}) \dodoi{10.1121/10.0019550}.
    
    \bibitem{Iwami2024}
    \bibinfo{author}{T.~Iwami} and \bibinfo{author}{A.~Omoto}, \enquote{\bibinfo{title}{Estimation of instantaneous sound intensity field using a dense microphone array}},  \bibinfo{journal}{Acoust. Sci. \& Tech.} \textbf{45}(2), \bibinfo{pages}{98--105} (\bibinfo{year}{2024}) \dodoi{10.1250/ast.e23.43}.
    
    \bibitem{Ogawa2007}
    \bibinfo{author}{H.~Ogawa} and \bibinfo{author}{A.~Hirabayashi}, \enquote{\bibinfo{title}{Sampling theorem with optimum noise suppression}},  \bibinfo{journal}{Sampling Theory in Signal and Image Processing} \textbf{6}, \bibinfo{pages}{167--184} (\bibinfo{year}{2007}).
    
    \bibitem{Shimamura2009}
    \bibinfo{author}{S.~Shimamura} and \bibinfo{author}{I.~Yamada}, \enquote{\bibinfo{title}{A robust function estimation in reproducing kernel hilbert space based on finite dimensional reformulations}}, in \emph{\bibinfo{booktitle}{Proc. APSIPA ASC 2009}} (\bibinfo{year}{2009}), pp. \bibinfo{pages}{426--429}.
    
    \bibitem{Hansen2010}
    \bibinfo{author}{P.~C. Hansen}, \emph{\bibinfo{title}{Discrete inverse problems: insight and algorithms}}  (\bibinfo{publisher}{SIAM}, \bibinfo{year}{2010}).
    
    \bibitem{Shawe2010}
    \bibinfo{author}{J.~Shawe-Taylor} and \bibinfo{author}{N.~Cristianini}, \emph{\bibinfo{title}{Kernel Methods for Pattern Analysis}}  (\bibinfo{publisher}{Cambridge University Press}, \bibinfo{year}{2004}).
    
    \bibitem{Boaz2015}
    \bibinfo{author}{B.~Rafaely}, \emph{\bibinfo{title}{Fundamentals of spherical array processing}}, Vol.~\bibinfo{volume}{8}  (\bibinfo{publisher}{Springer-Verlag}, \bibinfo{address}{Berlin}, \bibinfo{year}{2015}).
    
    \bibitem{Fliege1999}
    \bibinfo{author}{J.~Fliege} and \bibinfo{author}{U.~Maier}, \enquote{\bibinfo{title}{The distribution of points on the sphere and corresponding cubature formulae}},  \bibinfo{journal}{IMA Journal of Numerical Analysis} \textbf{19}(2), \bibinfo{pages}{317--334} (\bibinfo{year}{1999}).
    
    \bibitem{Lecomte2016}
    \bibinfo{author}{P.~Lecomte}, \bibinfo{author}{P.-A. Gauthier}, \bibinfo{author}{C.~Langrenne}, \bibinfo{author}{A.~Berry}, and \bibinfo{author}{A.~Garcia}, \enquote{\bibinfo{title}{A fifty-node lebedev grid and its applications to ambisonics}},  \bibinfo{journal}{J. Audio Eng. Soc.} \textbf{64}(11), \bibinfo{pages}{868--881} (\bibinfo{year}{2016}).
    
    \bibitem{Nikiforov1988}
    \bibinfo{author}{A.~F. Nikiforov} and \bibinfo{author}{V.~B. Uvarov}, \emph{\bibinfo{title}{Special functions of mathematical physics}}, Vol. \bibinfo{volume}{205}  (\bibinfo{publisher}{Springer}, \bibinfo{address}{Berlin}, \bibinfo{year}{1988}).
    
    \end{thebibliography}
\end{document}